\documentclass[conference]{IEEEtran}
\IEEEoverridecommandlockouts

\usepackage{cite}
\usepackage{amsmath,amssymb,amsfonts}
\usepackage{algorithmic}
\usepackage{graphicx}
\usepackage{textcomp}
\usepackage{xcolor}
\usepackage{booktabs}
\usepackage{subcaption}
\usepackage{amsthm}
\newtheorem{definition}{Definition}
\def\BibTeX{{\rm B\kern-.05em{\sc i\kern-.025em b}\kern-.08em
    T\kern-.1667em\lower.7ex\hbox{E}\kern-.125emX}}
\begin{document}

\title{Adaptive Lighting Control in Visible Light Systems: An Integrated Sensing, Communication, and Illumination Framework\\}

\author{
    \IEEEauthorblockN{Xinyan Xie\textsuperscript{$\dagger$}, 
    Xuesong Wang\textsuperscript{$\ddagger$}, 
    Xin Lai\textsuperscript{$\dagger$}, 
    Yongheng Wen\textsuperscript{$\dagger$}, 
    Fengrui Yang\textsuperscript{$\dagger$}, \\
    Haoyang He\textsuperscript{$\dagger$}, 
    Lai Zhang\textsuperscript{$\dagger$}, 
    Dong Zhao\textsuperscript{$* \dagger$}}
    
    \IEEEauthorblockA{\textsuperscript{$\dagger$}Fiber Optics Research Center, College of Smart Materials and Future Energy, Fudan University, Shanghai, China\\
    \textsuperscript{$\ddagger$}School of Science and Engineering, The Chinese University of Hong Kong, Shenzhen, Guangdong, China\\
    \textsuperscript{*}Email: zhaodong@fudan.edu.cn}
}

\maketitle

\begin{abstract}
Indoor visible light communication (VLC) is a promising sixth-generation (6G) technology, as its directional and sensitive optical signals are naturally suited for integrated sensing and communication (ISAC). However, current research mainly focuses on maximizing data rates and sensing accuracy, creating a conflict between high performance, high energy consumption, and user visual comfort. This paper proposes an adaptive integrated sensing, communication, and illumination (ISCI) framework that resolves this conflict by treating energy savings as a primary objective. The framework's mechanism first partitions the receiving plane using a geometric methodology, defining an activity area and a surrounding non-activity area to match distinct user requirements. User location, determined using non-line-of-sight (NLOS) sensing, then acts as a dynamic switch for the system's optimization objective. The system adaptively shifts between minimizing total transmit power while guaranteeing communication and illumination performance in the activity area and maximizing signal-to-noise ratio (SNR) uniformity in the non-activity area. Numerical results confirm that this adaptive ISCI approach achieves 53.59\% energy savings over a non-adaptive system and improves SNR uniformity by 57.79\%, while satisfying all illumination constraints and maintaining a mean localization error of 0.071~m.
\end{abstract}


\section{Introduction}
\label{sec:I}
Visible light communication (VLC) has emerged as a promising wireless technology that transmits data by modulating signals from LEDs. Taking advantage of the wide and unlicensed optical spectrum, VLC offers high data rates and is considered a key enabler to meet sixth-generation (6G) performance targets~\cite{2024-VLC-6G}. Meanwhile, integrated sensing and communication (ISAC) has become a fundamental research direction in 6G, aiming to unify communication and environmental sensing functions within a single platform~\cite{6G-ISAC}. Since optical signals are directional and sensitive to environmental changes, VLC can naturally support both data transmission and user location, making it well suited for indoor ISAC.

Current research on VLC-based ISAC systems mainly focuses on three directions: 1) advanced waveform design to balance communication and sensing performance~\cite{PAM,CAP}; 2) hardware integration using components like optical phased arrays (OPA) and reconfigurable intelligent surfaces (RIS)~\cite{OPA,LC-RIS,OIRS}; 3) advanced system architectures employing retroreflectors and angular diversity photodetectors~\cite{Retro-ISAC-1,AD-PD}. While most studies only consider line-of-sight (LOS) scenarios, recent studies have shown that non-line-of-sight (NLOS) channels can serve as valuable resources for communication and sensing, particularly in dynamic or obstructed scenarios~\cite{NLOS1,NLOS2}. However, the primary focus of these studies remains on improving sensing accuracy and communication data rates. Achieving such high performance often necessitates high transmit power, thereby leading to two significant practical problems: first, unnecessary energy consumption, especially when users are not performing focused work; and second, the visual discomfort caused by excessive illumination.

To address these problems, this paper proposes an adaptive integrated sensing, communication, and illumination (ISCI) framework that treats energy savings as a co-equal system objective. The central premise is that user requirements are not uniform. For example, focused work demands high signal-to-noise ratio (SNR) and illumination, while transit requires high SNR uniformity. Based on this, we first introduce a geometric methodology to partition the space into a high-performance activity area and a surrounding non-activity area to match these distinct needs. Then, motivated by~\cite{NLOS1,NLOS2}, we use NLOS sensing for real-time user localization, which acts as a dynamic switch for the system's optimization objective. The system shifts from minimizing power while guaranteeing high communication and illumination performance in the activity area to maximizing SNR uniformity in the non-activity area or entering a low-power sensing mode when no user is present. This adaptive approach thus achieves significant energy savings by matching power consumption to user requirements while ensuring reliable ISCI performance.

\section{System model}
\label{sec:II}
\begin{figure}[t]
\centerline{\includegraphics[width=1\linewidth]{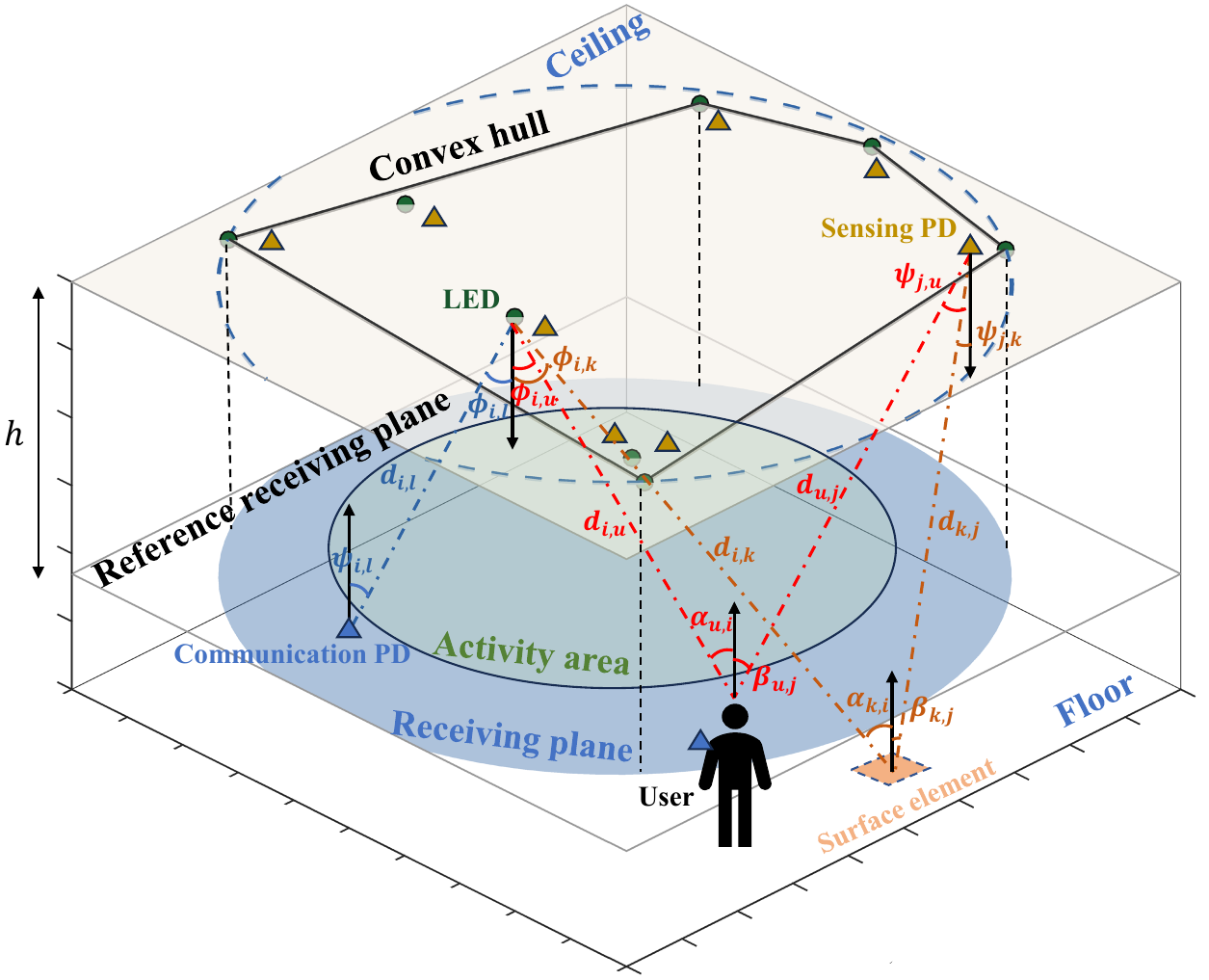}}
\caption{Schematic diagram of the system and channel model.}
\label{system_model}
\end{figure}

We consider an indoor space with $M$ LEDs and $N$ sensing photodiodes (PDs) located on the ceiling. The floor is discretized into $K$ surface elements, and the light emitted by the LEDs is reflected by these surface elements and then received by the sensing PDs. When a user enters the space, the user reflects incident light and alters the optical power received by the PDs. A reference receiving plane is located $h$ below the ceiling, and communication PDs can be distributed on it. This indoor VLC system involves two channel models: 1) a LOS communication/illuminance model between LEDs and communication PDs, and 2) a NLOS sensing model between LEDs and sensing PDs. In the following subsections, we present these models and detail the geometric construction method for our ISCI system. The overall system and channel model are shown in Fig.~\ref{system_model}.

\subsection{LOS Channel: Communication and Illumination Models}
\label{Channel_model}
Assume that each LED follows a generalized Lambertian radiation pattern with identical optical parameters. By neglecting multipath effects, the LOS channel gain between the $i$-th LED ($i=1,...,M$) and the $l$-th communication PD in the VLC system can be expressed as~\cite{Komine_2004}
\begin{equation}
\label{LOS_channel}
h_{L_{i,l}} = 
\begin{cases}
\begin{aligned}
&\frac{(m+1) A_c T_s(\psi_{l,i}) g(\psi_{l,i})}{2\pi d_{i,l}^2} \\
&\quad \times \cos^m(\phi_{i,l})\cos(\psi_{l,i}), 
\end{aligned}
& 0 \leq \psi_{l,i} \leq \Psi_c, \\
0, & \psi_{l,i} > \Psi_c,
\end{cases}
\end{equation}
where $A_c$ is the detector area of the communication PD, $d_{i,l}$ is the distance between the $i$-th LED and the $l$-th PD. The Lambertian order $m$ is calculated as $m = -\ln 2 / \ln \cos(\Phi_{1/2})$, where $\Phi_{1/2}$ is the semi-angle at half power.
$\phi_{i,l}$ is the irradiance angle, $\psi_{l,i}$ is the incidence angle, $T_s(\psi_{l,i})$ is the gain of an optical filter (i.e., typically $T_s(\cdot)=1$), and $ g(\psi_{l,i})$ is the gain of an optical concentrator, which is given as
\begin{equation}
g(\psi_{l,i}) =
\begin{cases}
\displaystyle
\frac{n^2}{\sin^2(\Psi_c)}, & 0 \leq \psi_{l,i} \leq \Psi_c, \\
0, & \psi_{l,i} > \Psi_c,
\end{cases}
\end{equation}
where $n$ is the refractive index of the $l$-th PD and $\Psi_c$ is the width of the field-of-vision (FOV) at a communication PD. Similarly, the illuminance at the $l$-th PD can be expressed as~\cite{komine_2009}
\begin{equation}
\label{illumination}
e_{i,l} = \frac{I_i(0)}{d_{i,l}^2} \cos^m(\phi_{i,l})\cos(\psi_{l,i}),
\end{equation}
where $I_i(0) = \frac{m+1}{2\pi}\eta P_{t_i}$ is the center radiant intensity of the $i$-th LED, with luminous efficacy $\eta$ and optical power $P_{t_i}$.

\subsection{NLOS Channel: Sensing Models}
\label{sec:NLOS Channel}
When no user is present in the space at time $t_0$, the NLOS channel gain with first-order reflections between the $i$-th LED, the $k$-th surface element ($k=1,...,K$), and the $j$-th sensing PD ($j=1,...,N$) can be expressed as~\cite{NLOS1}
\begin{equation}
\label{NLOS_channel}
\begin{split}
h_{N_{i,k,j}}^{(t_0)} &= \frac{\rho_{k}(m+1) A_s A_k}{2\pi^2 d_{i,k}^2 d_{k,j}^2} \cos^m(\phi_{i,k}) \cos(\psi_{j,k}) \\
&\quad \times \cos(\alpha_{k,i}) \cos(\beta_{k,j}) T_s(\psi_{j,k}) g(\psi_{j,k}),
\end{split}
\end{equation}
where $\rho_k$ is the reflection coefficient of the $k$-th surface element, and $A_k$ is the area of the $k$-th surface element. The other parameters in Eq.~\eqref{NLOS_channel} are defined similarly to their counterparts in Eq.~\eqref{LOS_channel}.

When a user enters the space at time $t_u$, the channel gain still follows the Lambertian diffuse reflection model. The user acts as a surface element, and the reflection path is established between the $i$-th LED, the $u$-th user, and the $j$-th PD. The NLOS channel gain is similar to Eq.~\eqref{NLOS_channel}, with the surface element replaced by the user~\cite{NLOS1}
\begin{equation}
\label{user_channel}
\begin{split}
h_{N_{i,u,j}}^{(t_u)} &= \frac{\rho_{u}(m+1) A_s A_u}{2\pi^2 d_{i,u}^2 d_{u,j}^2} \cos^m(\phi_{i,u}) \cos(\psi_{j,u}) \\
&\quad \times \cos(\alpha_{u,i}) \cos(\beta_{u,j}) T_s(\psi_{j,u}) g(\psi_{j,u}),
\end{split}
\end{equation}
where $\rho_u$ and $A_u$ are the reflection coefficient and the area of the $u$-th user. Note that for Eqs.~\eqref{NLOS_channel} and~\eqref{user_channel}, the channel gain is non-zero only when the incident angle falls within the receiver's FOV (i.e., $0 \le \psi_{j,k}/\psi_{j,u} \le \Psi_s$).

When no user is present in the space, the optical power received by the $j$-th PD is the sum of contributions from all LEDs via all surface elements, and the total received power can be expressed as
\begin{equation}
P_{r_j}^{(t_0)} = \sum_{i=1}^{M} \sum_{k=1}^{K} P_{t_i}^{(t_0)} h_{N_{i,k,j}}^{(t_0)}.
\end{equation}

When a user enters the space, each PD receives additional optical power due to light reflected from the user's body. The total received power at the $j$-th PD is given by
\begin{equation}
P_{r_j}^{(t_u)} = \sum_{i=1}^{M} P_{t_i}^{(t_u)} \left( \sum_{k \in \mathcal{K}} h_{N_{i,k,j}}^{(t_u)} + \sum_{u \in \mathcal{U}} h_{N_{i,u,j}}^{(t_u)} \right),
\end{equation}
where $\mathcal{K}$ denotes the set of surface elements that are not occluded by the user at time $t_u$, and $\mathcal{U}$ denotes the set of reflective elements on the user's body. Therefore, the presence of a user in the space leads to a variation in the received power at each sensing PD, compared to the no-user scenario. This variation at time $t_u$ can be expressed as
\begin{equation}
\label{delta_P}
\Delta P_{r_j}^{(t_u)} = \left| P_{r_j}^{(t_u)} - P_{r_j}^{(t_0)} \right|.
\end{equation}

According to Eq.~\eqref{delta_P}, the user sensing process relies on the similarity between the actual and predicted power variations at each sensing PD. It comprises two phases: an offline phase for precomputing a reference dataset and an online phase for real-time position estimation. In the offline phase, the $K$ candidate positions $(x_k, y_k)$ are defined by the centers of the $K$ surface elements discretized previously. For each position, the expected power variation at each sensing PD is simulated under the assumption that a user is located at that position. The predicted values, denoted by $\Delta P_{r_j,\text{predict}}^{(t_u)}(x_k, y_k)$, are computed for all PDs and stored in a lookup table indexed by $k$.

In the online phase, the actual power variation $\Delta P_{r_j,\text{actual}}^{(t_u)}$ is measured. The localization loss $L_k$ is then calculated as the mean squared error (MSE) between actual and predicted power variations, given by
\begin{equation}
L_k = \sum_{j=1}^{N} \left( \Delta P_{r_j,\text{actual}}^{(t_u)} - \Delta P_{r_j,\text{predict}}^{(t_u)}(x_k, y_k) \right)^2.
\end{equation}
The estimated position $(x^*, y^*)$ is the candidate $(x_{k^*}, y_{k^*})$ that minimizes this loss, given by
\begin{equation}
(x^*, y^*) = (x_{k^*}, y_{k^*}), \quad k^* = \arg\min_{k} L_k.
\end{equation}

\subsection{Geometric Construction Methodology}
\label{sec:IIC}
According to Eqs.~\eqref{LOS_channel} and~\eqref{illumination}, the received power and illuminance are related to the distance from the LEDs. PDs at spatial boundaries suffer from poor performance, and these areas are also rarely occupied in realistic user mobility models~\cite{RWP}. Given this combination of poor performance and low user activity, it is inefficient to include these boundary regions in the system optimization. We therefore first identify the high-performance core region. This region is defined by the convex hull $\mathcal{C}$ of the 2D LED projections, $\mathcal{S} = \{s_i \mid i=1,\ldots,M\}$:
\begin{equation}
    \mathcal{C}=conv(\mathcal{S}).
\end{equation}

While PDs within $\mathcal{C}$ achieve relatively high performance due to their proximity to the LEDs, this convex hull is unsuitable as the new receiving plane because its potentially irregular geometry is computationally complex and difficult to analyze. Our goal is to define a new plane that is geometrically regular, computationally tractable, and uniquely determined for any given LED deployment. The minimum enclosing circle (MEC) of the convex hull $\mathcal{C}$ is the ideal geometric construct that satisfies these requirements, as it is the smallest regular circle that contains the entire high-performance core region. We therefore define the new receiving plane as follows.
\begin{definition}[Receiving Plane Definition]
\label{Def:receiving_plane}
Let $\mathcal{S}_{\mathrm{MEC}}$ be the minimum enclosing circle (MEC) of the convex hull $\mathcal{C}$, and let $\mathcal{B}$ denote the set of points within the physical boundaries of the space. The new receiving plane $\mathcal{I}$ is defined as the intersection of these two sets:
\begin{equation}
\mathcal{I} = \mathcal{S}_{\mathrm{MEC}} \cap \mathcal{B}.
\end{equation}
\end{definition}

However, user behavior across this plane is not uniform. To realistically model this, we adopt the principles of the ISO/CIE 8995-1 standard, which links illumination targets to distinct user activities (e.g., activity, surrounding and background). We therefore partition $\mathcal{I}$ into a high-performance activity area $\mathcal{W}$, intended for focused user work, and a surrounding non-activity area. The definition for $\mathcal{W}$ must satisfy two criteria: it supports the highest communication and illumination demands, and it needs to be geometrically regular for computational tractability. The maximum inscribed circle (MIC) of the original convex hull $\mathcal{C}$ is the ideal geometric construct that uniquely satisfies both requirements. We thus define the activity area $\mathcal{W}$ as the region enclosed by the MIC, as illustrated in Fig.~\ref{system_model}.

\textbf{Remark:} \textit{Unless specified otherwise, the receiving plane in the following sections refers to the MEC-defined region $\mathcal{I}$.}

\section{ISCI Framework: Lighting Control Strategy}
\label{sec:III}
As described in Section~\ref{sec:NLOS Channel}, user localization is performed by computing the MSE between actual and predicted power variations. Once the user's position is determined, the corresponding ISCI framework is applied based on user behavior, aiming to reduce energy consumption while maintaining user demands. Specifically, three ISCI modes are defined:
\begin{itemize}
\item \textbf{No-user mode}: user absent from the receiving plane;
\item \textbf{Uniformity mode}: user located in the non-activity area;
\item \textbf{Enhanced mode}: user located in the activity area.
\end{itemize}

\subsection{No-user Mode}
When no user is present in the receiving plane, communication and illumination are not needed, and only the sensing requirement needs to be considered. Thus, all LEDs operate at the minimum power level $P_{\min}$ to reduce energy consumption and continue sensing.

\subsection{Uniformity Mode}
When the user is present in the non-activity area, their behavior typically involves movement and short periods of stillness. This area does not require high-performance communication or illumination, since the primary goal is to provide a stable service as the user moves. To reduce communication signal fluctuations, the uniformity mode adjusts the transmit power of the LEDs to improve SNR uniformity across the receiving plane. The SNR at the $l$-th PD is given by
\begin{equation}
\label{SNR}
\Lambda_l = \frac{(R P_{r_l})^2}{\sigma_{\text{shot}}^2 + \sigma_{\text{thermal}}^2},
\end{equation}
where $R$ denotes the PD responsivity, $P_{r_l}=\sum_{i=1}^{M} h_{L_{i,l}} P_{t_i}$ is the total received power at the $l$-th PD. $\sigma_{\text{shot}}^2$ is the shot noise variance, 
and $\sigma_{\text{thermal}}^2$ is the thermal noise variance, given by

\begin{equation}
\label{shot}
\sigma_{\text{shot}}^2 = 2q R P_{r_l} B + 2q I_B I_2 B,
\end{equation}

\begin{equation}
\label{thermal}
\sigma_{\text{thermal}}^2 = \frac{8\pi k T_k}{G_{\text{ol}}} C_{\text{pd}} A I_2 B^2 
+ \frac{16\pi^2 k T_k \Gamma}{g_m} C_{\text{pd}}^2 A^2 I_3 B^3.
\end{equation}
The definitions for the various parameters in Eqs.~\eqref{shot} and~\eqref{thermal} are adopted from~\cite{Komine_2004}. To simplify the model, we define a specific scenario based on high-speed VLC with reliable channel conditions. This involves the following two simplifications.
\begin{itemize}
\item \textbf{Optical parameter simplification}: the communication PD's FOV is set to $\Psi_c = 90^\circ$ and the LED's half-power angle to $\Phi_{1/2} = 60^\circ$ (i.e., Lambertian order $m = 1$);
\item \textbf{Noise simplification}: assuming a high-SNR reliable channel, the constant noise terms $\left(2q I_B I_2 B + \sigma_{\text{thermal}}^2\right)$ in Eqs.~\eqref{SNR}--\eqref{thermal} are neglected. The simplified SNR can be expressed as $\Lambda_l = R / (2qB) \cdot P_{r_l}$.
\end{itemize}

Based on these simplifications, improving SNR uniformity is equivalent to minimizing the SNR variance across the receiving plane, expressed as
\begin{equation} 
\label{SNR_Var}
\min_{P_{t_i}} \quad \frac{1}{L} \sum_{l=1}^{L} \left( \Lambda_l - \bar{\Lambda} \right)^2,
\end{equation} 
where $\bar{\Lambda}=\frac{1}{L} \sum_{l=1}^{L} \Lambda_l$ is the average SNR of PDs ($l=1,...,L$) over the receiving plane. By substituting Eq.~\eqref{SNR} into Eq.~\eqref{SNR_Var}, the objective function can be expressed as
\begin{equation} 
\label{obj_final} 
\min_{P_{t_i}} \quad \frac{1}{L} \sum_{l=1}^{L} \left( \frac{R A_c h^2 n^2}{2 q \pi B} \sum_{i=1}^{M} \frac{P_{t_i}}{d_{i,l}^4} - \frac{1}{L} \sum_{l=1}^{L} \Lambda_l \right)^2. 
\end{equation}

To reformulate this problem into a standard quadratic program (QP), let $C = \frac{R A_c h^2 n^2}{2q\pi B}$, and define the LED power vector as $\mathbf{p} = \left[P_{t_1}, \dots, P_{t_M}\right]^\top \in \mathbb{R}^{M \times 1}$. The SNR at the $l$-th PD can be expressed as $\Lambda_l = \mathbf{a}_l^\top \mathbf{p}$, with $\mathbf{a}_l = \left[ {\frac{C}{d_{1l}^4}}, \dots, {\frac{C}{d_{Ml}^4}} \right]^\top$. Let $\mathbf{A} \in \mathbb{R}^{L \times M}$ be the matrix with rows $\mathbf{a}_l^\top$, the average SNR can be expressed as
\begin{equation}
\bar{\Lambda} = \frac{1}{L} \mathbf{1}^\top \mathbf{A} \mathbf{p}.
\end{equation}
Then the SNR variance can be given by
\begin{equation}
\label{A:SNR}
\frac{1}{L} \sum_{l=1}^{L} (\Lambda_l - \bar{\Lambda})^2 = \frac{1}{L} \left\| \mathbf{A} \mathbf{p} - \frac{1}{L} \mathbf{1} \cdot \mathbf{1}^\top \mathbf{A} \mathbf{p} \right\|^2.
\end{equation}

Define $ \mathbf{M} = \mathbf{I} - \frac{1}{L} \mathbf{1} \cdot \mathbf{1}^\top$, Eq.~\eqref{A:SNR} simplifies to
\begin{equation}
\frac{1}{L} \left\| \mathbf{M} \mathbf{A} \mathbf{p} \right\|^2 
= \frac{1}{L} \mathbf{p}^\top \mathbf{A}^\top \mathbf{M} \mathbf{A} \mathbf{p},
\end{equation}
and define $\mathbf{Q} = \frac{1}{L} \mathbf{A}^\top \mathbf{M} \mathbf{A}$, the objective function becomes
\begin{equation}
\label{QP}
\min_{\mathbf{p}} \ \mathbf{p}^\top \mathbf{Q} \mathbf{p}.
\end{equation}

To ensure visual comfort and system feasibility, the following constraints define the acceptable illuminance range for indoor work, as well as the LED power limits required to protect the LED lifespan and maintain sensing capabilities.
\begin{subequations}
\begin{align}
\text{s.t.} \quad
E_{\min_u} &\leq E_{r_{l}}=\sum_{i=1}^{M} e_{i,l} \leq E_{\max_u}, \quad \forall l \in \{1, 2, \dots, L\}, \label{const_a} \\
P_{\min} &\leq P_{t_i} \leq P_{\max}, \quad \forall i \in \{1, 2, \dots, M\}. \label{const_b}
\end{align}
\end{subequations}
The positive semi-definiteness of $\mathbf{Q}$ ensures the convexity of the quadratic objective function in Eq.~\eqref{QP}, and all constraints introduced in the uniformity mode are linear, resulting in a convex feasible set. Therefore, the optimization problem is a standard convex QP and can be efficiently solved using standard QP solvers.

\begin{figure*}[t]
    \centering
    
    \begin{subfigure}[t]{0.30\textwidth}
        \centering
        \includegraphics[width=\linewidth]{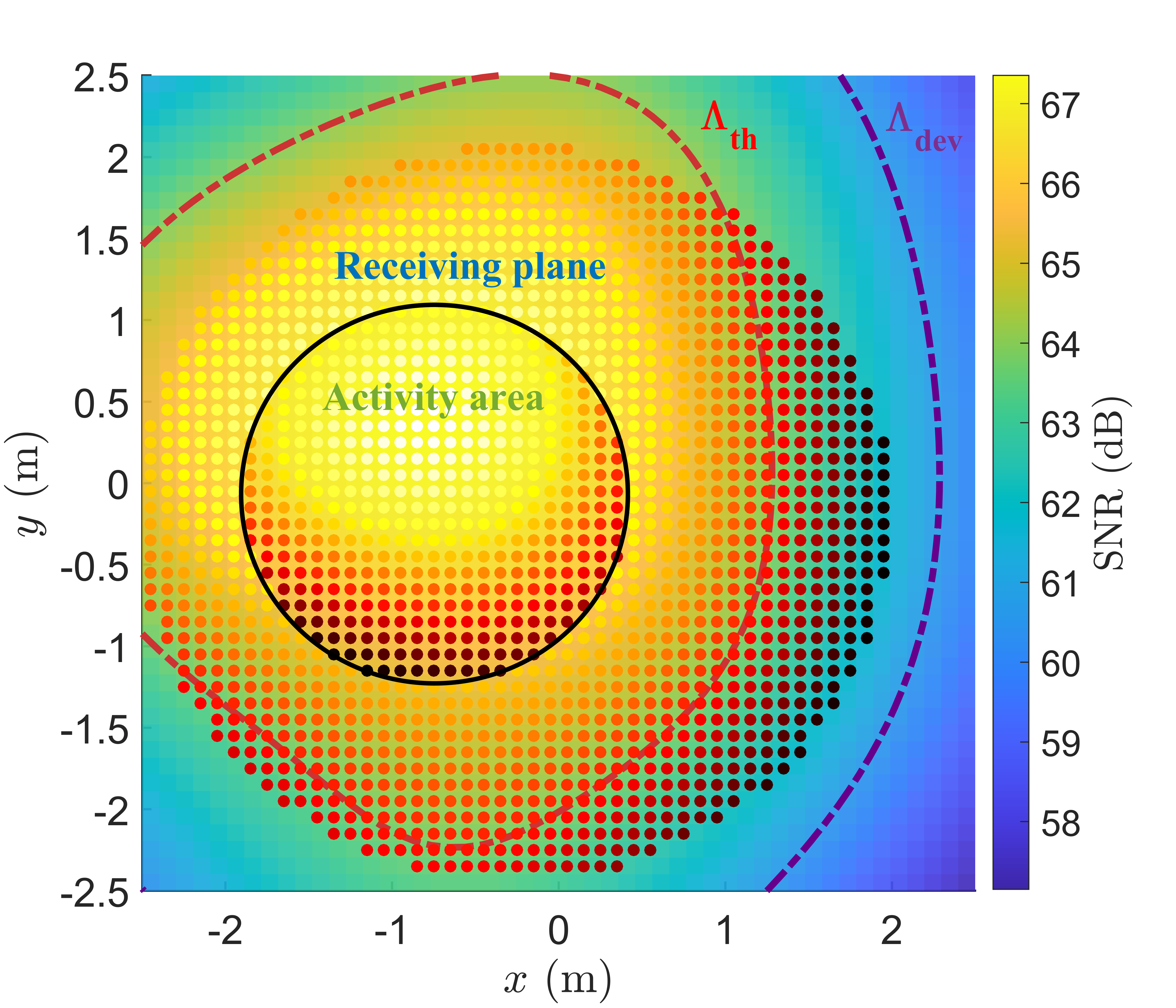}
        \caption{}
        \label{fig:sub-a}
    \end{subfigure}
    \hfill
    \begin{subfigure}[t]{0.30\textwidth}
        \centering
        \includegraphics[width=\linewidth]{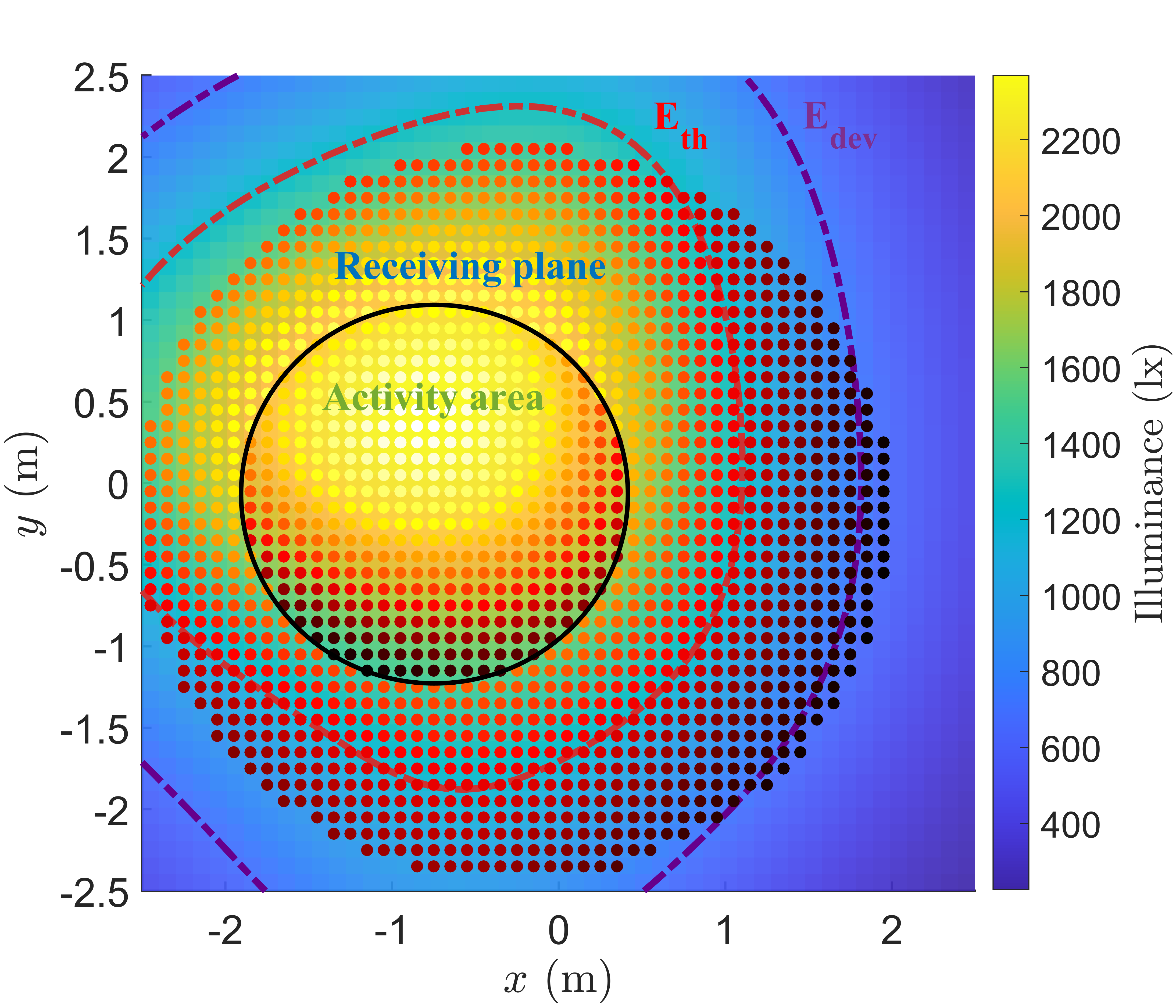}
        \caption{}
        \label{fig:sub-b}
    \end{subfigure}
    \hfill
    \begin{subfigure}[t]{0.30\textwidth}
        \centering
        \includegraphics[width=\linewidth]{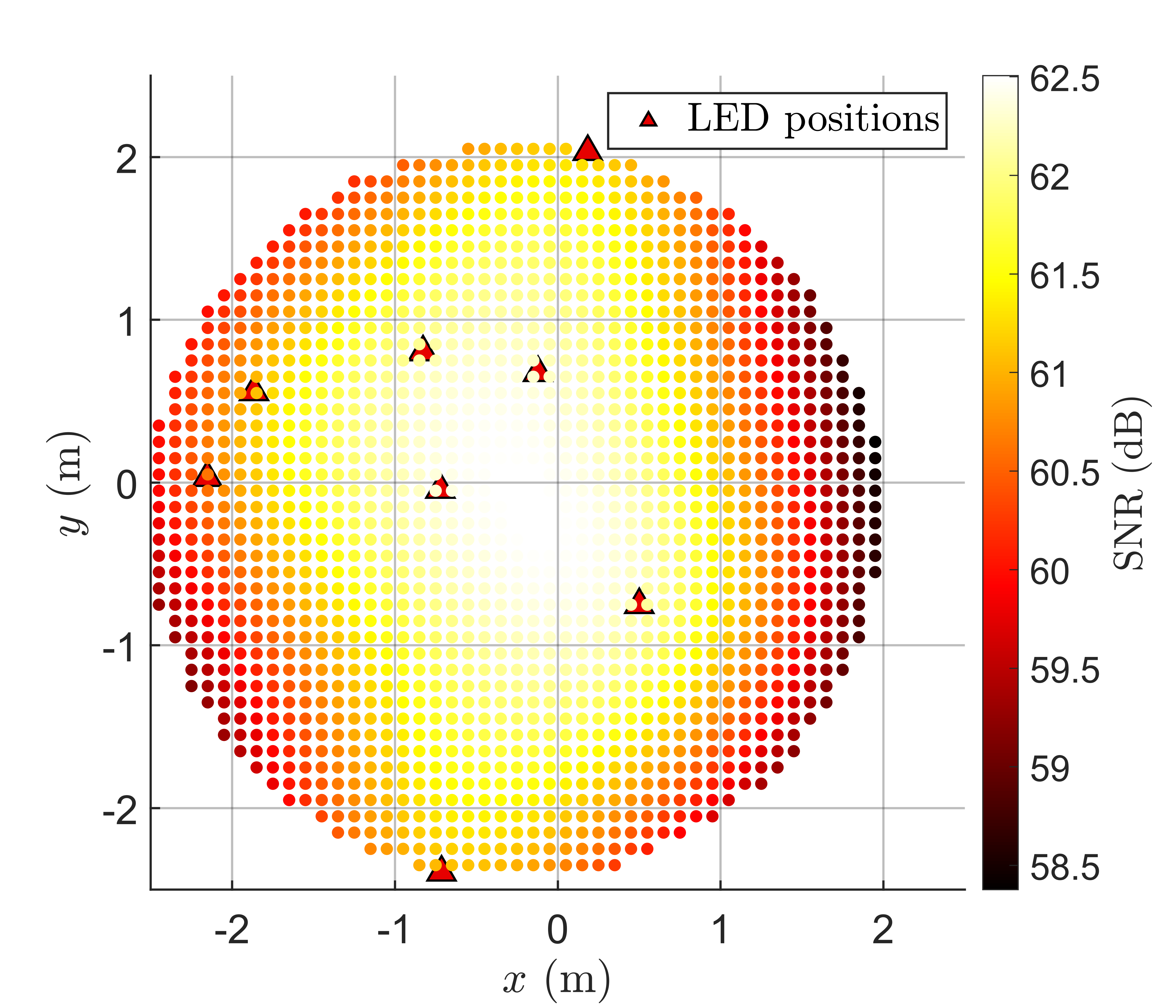}
        \caption{}
        \label{fig:sub-c}
    \end{subfigure}
    
    \vspace{1mm} 

    \begin{subfigure}[t]{0.30\textwidth}
        \centering
        \includegraphics[width=\linewidth]{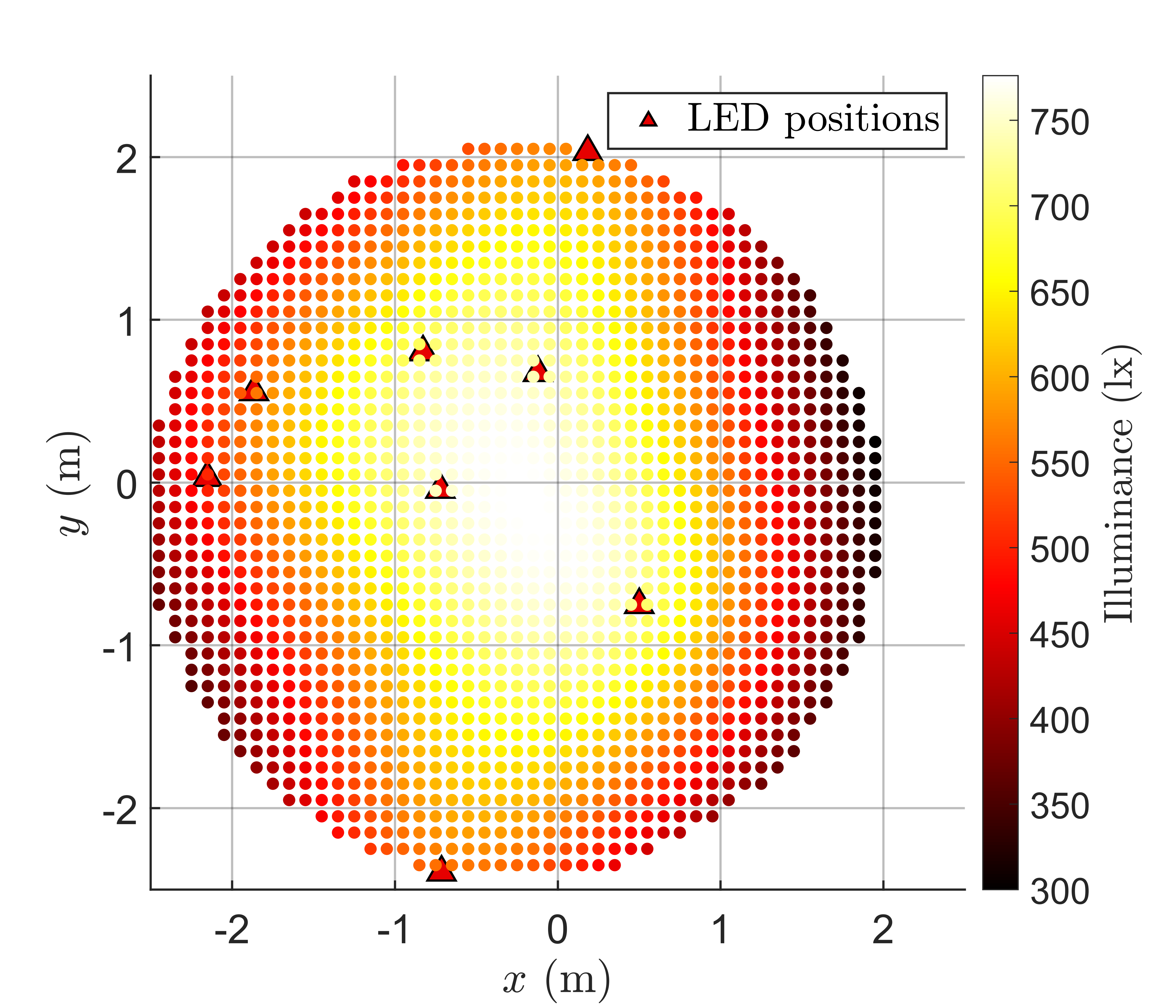}
        \caption{}
        \label{fig:sub-d}
    \end{subfigure}
    \hfill
    \begin{subfigure}[t]{0.30\textwidth}
        \centering
        \includegraphics[width=\linewidth]{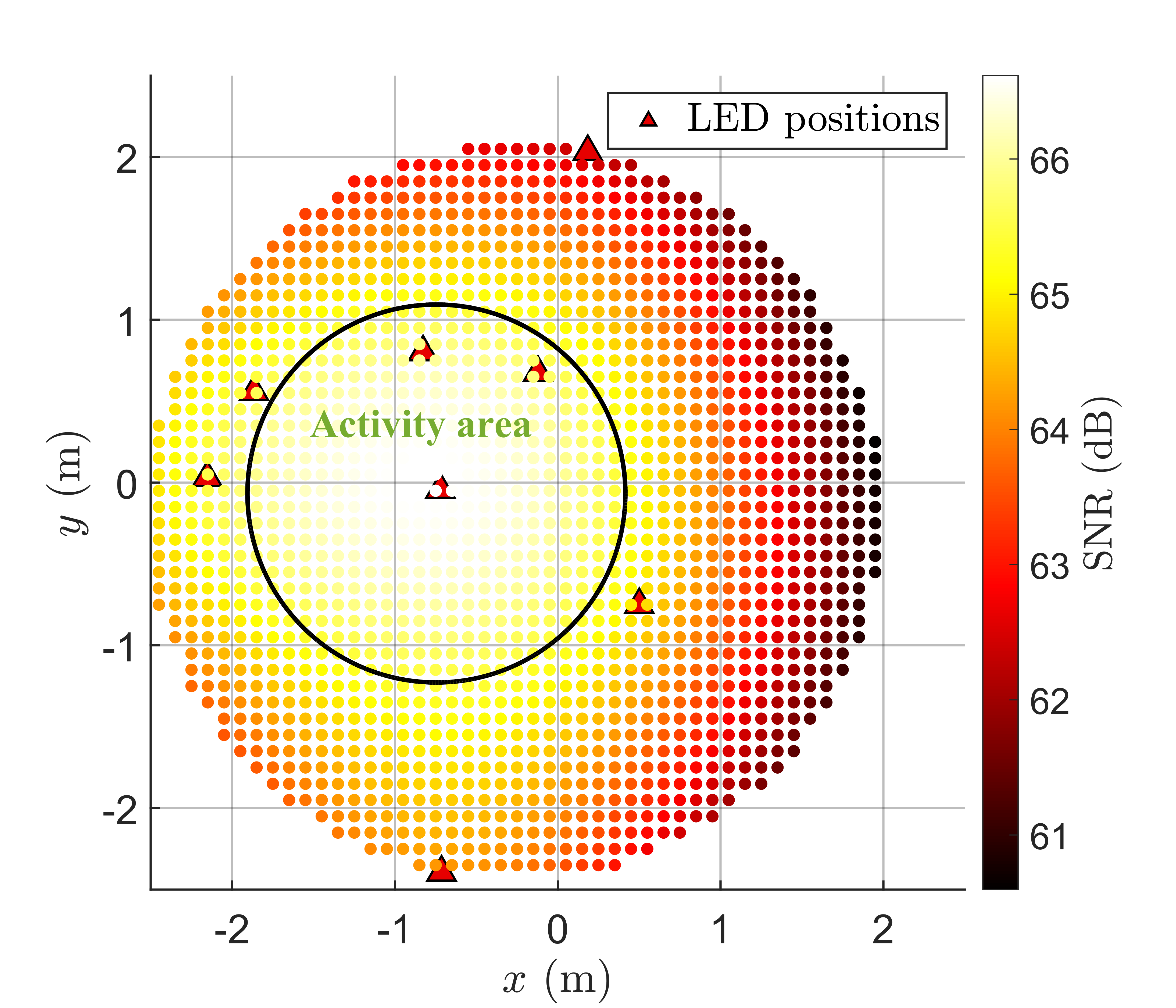}
        \caption{}
        \label{fig:sub-e}
    \end{subfigure}
    \hfill
    \begin{subfigure}[t]{0.30\textwidth}
        \centering
        \includegraphics[width=\linewidth]{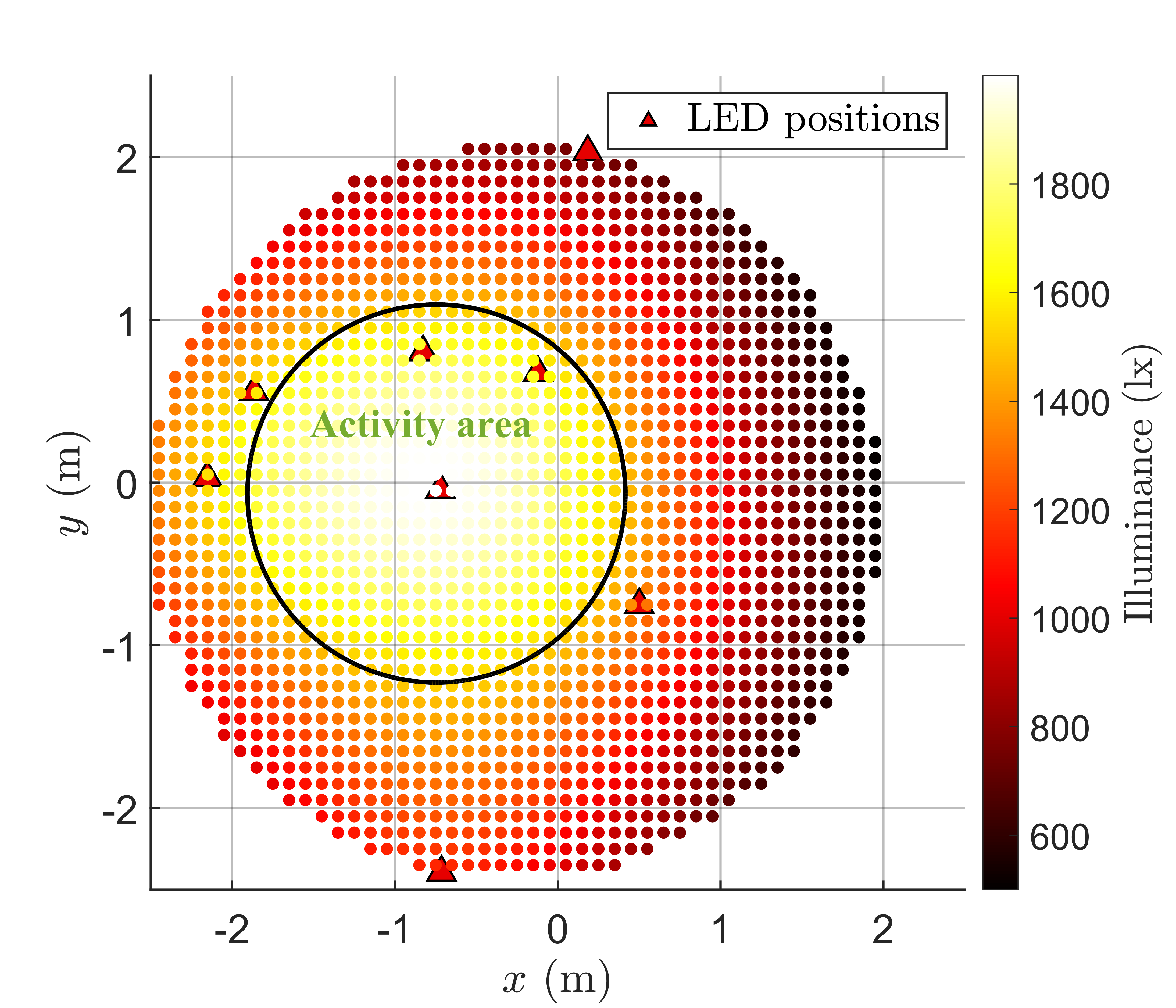}
        \caption{}
        \label{fig:sub-f}
    \end{subfigure}

    \caption{Distribution of SNR and illuminance. (a) Baseline SNR and (b) illuminance distribution; (c) SNR and (d) illuminance under the uniformity mode; (e) SNR and (f) illuminance under the enhanced mode.}
    \label{fig:2}
\end{figure*}

\subsection{Enhanced Mode}
When a user is detected in the activity area, the ISCI system switches to enhanced mode. In this mode, it is required that all locations within this area $\mathcal{W}$ satisfy two conditions: the SNR must exceed a predefined threshold $\Lambda_{\text{th}}$ for high-speed communications, and the illuminance must fall within a new enhanced range $[E_{\min_e}, E_{\max_e}]$. This ensures both reliable communication and adequate illumination throughout the entire activity area. On this basis, the objective is to minimize the total transmit power of the LEDs for energy savings. The optimization problem can be expressed as
\begin{equation}
\label{enhanced_mode}
\min_{P_{t_i}} \quad \sum_{i=1}^{M} P_{t_i},
\end{equation}
\begin{subequations}
\label{enhanced_constraint}
\begin{align}
\text{s.t.} \quad P_{\min} &\leq P_{t_i} \leq P_{\max}, && \forall i \in \{1, 2, \dots, M\}, \label{eq:con_power} \\
E_{\min_{e}} &\leq E_{r_{l}} \leq E_{\max_{e}}, && \forall l \in \mathcal{W}, \label{eq:con_illum} \\
\Lambda_{l} &\geq \Lambda_{\text{th}}, && \forall l \in \mathcal{W}. \label{eq:con_snr}
\end{align}
\end{subequations}

Eqs.~\eqref{enhanced_mode} and~\eqref{enhanced_constraint} constitute a standard linear program (LP) and can be efficiently solved using LP solvers.

\section{Numerical Results}
\label{sec:IV}
We consider an indoor space with dimensions $(5 \times 5 \times 3)~\text{m}^3$. The receiving plane is located at a vertical distance of $h = 2.15~\text{m}$ below the ceiling. A total of $M = 8$ LEDs are randomly deployed on the ceiling, with $N = 8$ sensing PDs located near their respective LEDs. Following the geometric construction methodology, the convex hull $\mathcal{C}$ is formed by the 2D projections of these LEDs. The activity area $\mathcal{W}$ is then uniquely determined as the MIC of this convex hull. Other system parameters follow~\cite{NLOS1,komine_2009}.

\subsection{Validation of Geometrically Defined Regions}
To quantitatively validate the performance of the MEC and MIC regions, we establish two SNR-based benchmarks: the average threshold $\Lambda_{\text{avg}}$ and the deviation threshold $\Lambda_{\text{dev}}$. These are formulated as
\begin{equation} 
\Lambda_{\text{avg}}  = \frac{1}{L_{\text{ref}}} \sum_{l=1}^{L_{\text{ref}}} \Lambda_l,
\end{equation}
\begin{equation} 
\Lambda_{\text{dev}} = \Lambda_{\text{avg}} - \frac{\Lambda_{\text{avg}} - \Lambda_{\text{min}}}{2},
\end{equation}
where $L_{\text{ref}}$ denotes the total number of communication PDs on the reference receiving plane, and $\Lambda_{\text{avg}}$ represents the average SNR across these PDs. The term $\Lambda_{\text{min}} = \min_{l} \Lambda_l$ denotes the minimum SNR observed on the reference plane. Consequently, $\Lambda_{\text{dev}}$ serves as a uniformity lower bound, defined mathematically as the midpoint between the average ($\Lambda_{\text{avg}}$) and the minimum ($\Lambda_{\text{min}}$) SNR values. Based on these thresholds, we define two evaluation criteria: a communication PD is considered to meet the high-performance requirement if $\Lambda_l > \Lambda_{\text{avg}}$, whereas it is considered to violate the uniformity requirement if $\Lambda_l < \Lambda_{\text{dev}}$.

As illustrated in Fig.~\ref{fig:2}(a), the red and purple lines denote the boundaries corresponding to these two thresholds. Compared to the reference receiving plane, the MEC receiving plane demonstrates significant improvement in satisfying these criteria. Specifically, on the reference plane, only 57.85\% of the PDs exceed the average threshold $\Lambda_{\text{avg}}$, while 10.91\% fall below the deviation threshold $\Lambda_{\text{dev}}$. In contrast, on the MEC plane, 79.31\% of the PDs exceed $\Lambda_{\text{avg}}$, and none violate the uniformity threshold (i.e., 0.00\% fall below $\Lambda_{\text{dev}}$). In addition, the SNR variance of the MEC receiving plane is 1.99. Furthermore, the MIC activity area achieves optimal performance, with 100\% of the PDs satisfying $\Lambda_j > \Lambda_{\text{avg}}$ and 0.00\% falling below $\Lambda_{\text{dev}}$. Similarly, the
 illuminance criteria $E_{\text{avg}}$ and $E_{\text{dev}}$ are defined in the same way
 as the SNR benchmarks, and the results are shown in Fig.~\ref{fig:2}(b).

\subsection{Performance of Uniformity and Enhanced Modes}
Different modes are associated with different optimization objectives and constraint sets, as described in Section~\ref{sec:III}. The transmit power for each LED is constrained to $[P_{\min}, P_{\max}] = [10~\text{W}, 80~\text{W}]$ to maintain sensing and protect the LED lifespan. Moreover, we set $E_{\min_u}$ and $E_{\max_u}$ to 300~lx and 1500~lx, respectively, in accordance with the recommended illuminance range for office work specified by the ISO standard~\cite{komine_2009}. The enhanced illuminance range $[E_{\min_e}, E_{\max_e}]$ is set to 500~lx above $[E_{\min_u}, E_{\max_u}]$ to support focused work or tasks requiring higher visual precision.

As shown in Fig.~\ref{fig:2}(c), in the uniformity mode, the SNR variance of the receiving plane is reduced to 0.84. Compared with the unoptimized results, the SNR uniformity is improved by 57.79\%. In addition, as shown in Fig.~\ref{fig:2}(d), the illuminance ranges from 300.00~lx to 776.15~lx, satisfying illumination constraints. Similarly, in the enhanced mode, the highest SNR values are concentrated in the activity area, as shown in Fig.~\ref{fig:2}(e), and all satisfy the constraints of Eq.~\eqref{eq:con_snr}. Illuminance ranges from 1502.87~lx to 2000~lx, satisfying the constraints of Eq.~\eqref{eq:con_illum}, as shown in Fig.~\ref{fig:2}(f). Meanwhile, the optimized total power is 300.06~W, a reduction compared to the unoptimized total power of 361.60~W. In addition, the LED power allocations under both modes comply with the power constraints specified in Eq.~\eqref{const_b} and Eq.~\eqref{eq:con_power}, while contributing to a reduction in overall energy consumption.

\begin{figure}[t]
    \centering
    \includegraphics[width=1\linewidth]{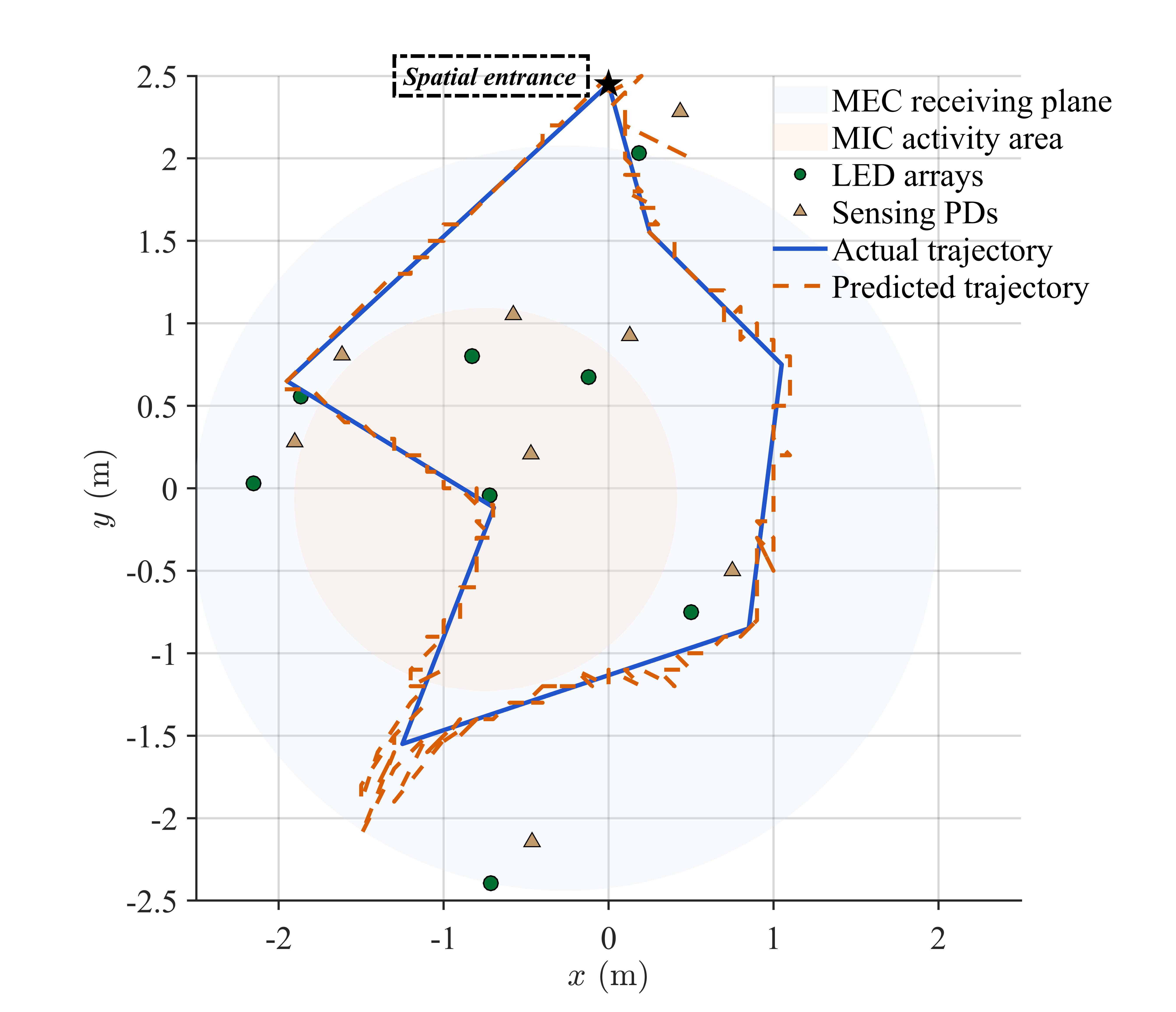}
    \caption{Actual and predicted user trajectories.}
    \label{fig:3}
\end{figure}

\begin{figure}[t]
    \centering
    \includegraphics[width=1\linewidth]{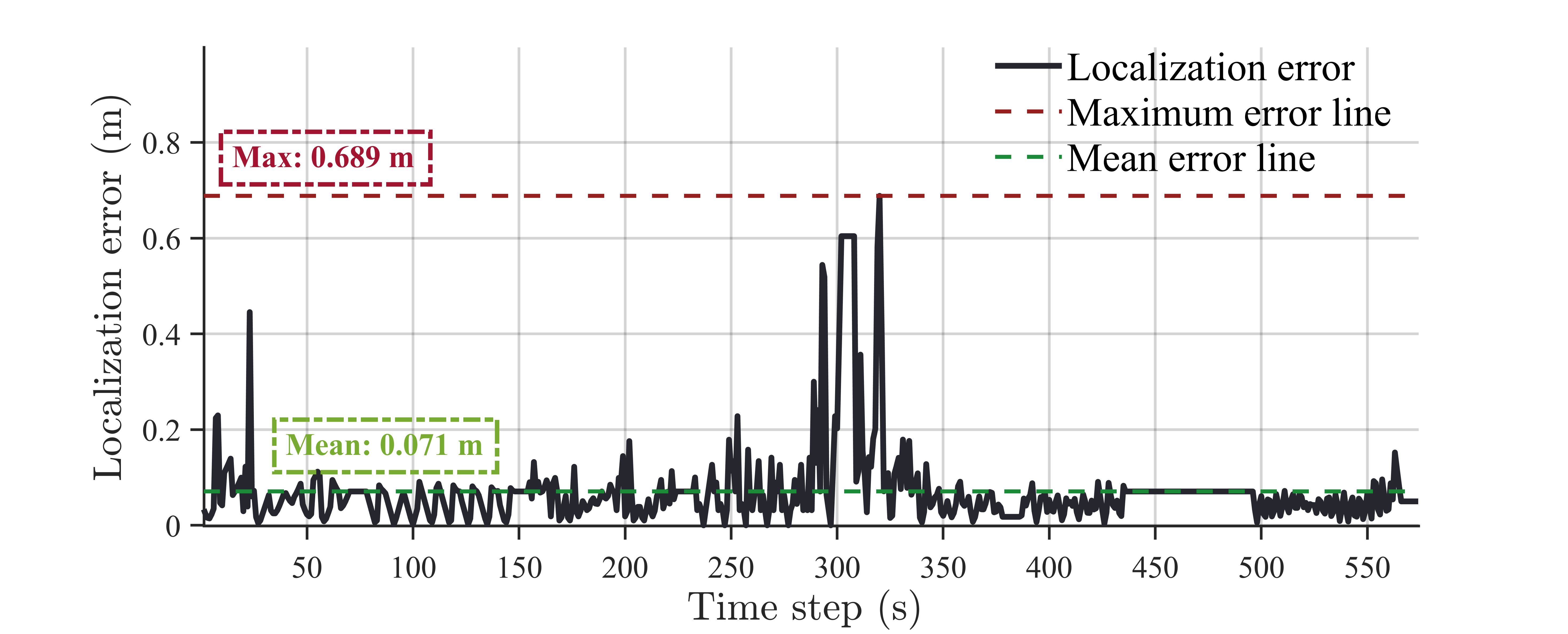}
    \caption{Variation of localization error.}
    \label{fig:4}
\end{figure}

\subsection{Localization Performance and Overall Energy Savings}
In our proposed ISCI framework, the sensing model directly determines the prediction of user location, thereby enabling accurate mode switching. To evaluate the localization performance, a representative user trajectory is designed as follows: the user first enters the space and moves randomly in the non-activity area of the MEC receiving plane. The user then moves into the activity area and remains stationary for a period to simulate focused work. Finally, the user exits the activity area, moves back through the non-activity area, and leaves the space. This is depicted in Fig.~\ref{fig:3}, where the blue solid line represents the user's actual trajectory, and the red dashed line indicates the predicted trajectory. Throughout the entire process, from entry to exit, the proposed ISCI framework achieves energy savings of 53.59\% compared to a non-adaptive VLC system (i.e., all LEDs operating at fixed normal power).

Furthermore, Fig.~\ref{fig:4} illustrates the localization error over time, corresponding to the actual and predicted trajectories shown in Fig.~\ref{fig:3}. The results show that the maximum localization error is 0.689~m, while the average error remains at a low level of 0.071~m. Notably, these results are achieved using a non-optimized, random LED configuration, suggesting that a strategic deployment of LEDs and sensing PDs could further enhance localization accuracy.

\section{Conclusion}
This paper presents an adaptive ISCI framework that addresses the conflict between ISCI performance and energy consumption in indoor VLC systems. The framework's core mechanism is the dynamic switching of its optimization objective based on user location. It shifts from minimizing total transmit power in the activity area to maximizing SNR uniformity in the non-activity area. This ISCI approach was validated by numerical results. The framework achieved 53.59\% energy savings over a non-adaptive system and improved SNR uniformity by 57.79\%, while simultaneously satisfying illumination constraints and maintaining a 0.071~m mean localization error. These results confirm the framework's ability to jointly manage ISCI performance and energy savings. Future work will extend the framework to complex multi-user environments to enhance robustness and investigate transition algorithms to mitigate communication interruptions and flicker.

\section*{Acknowledgment}
This work was supported by the Science and Technology Commission of Shanghai Municipality (17DZ2280600).

\bibliographystyle{IEEEtran}
\bibliography{References}

\vspace{12pt}

\end{document}